\documentclass[5p,twocolumn, 11pt]{elsarticle}
\makeatletter
\def\ps@pprintTitle{%
 \let\@oddhead\@empty
 \let\@evenhead\@empty
 \def\@oddfoot{}%
 \let\@evenfoot\@oddfoot}
\makeatother
\usepackage[mathscr]{euscript}
\usepackage{enumitem} 

\usepackage{dblfloatfix}
\usepackage{float}
\usepackage{graphicx}
\usepackage{caption}
\usepackage{subcaption} 

\usepackage{dblfloatfix}
\usepackage{wrapfig}
\usepackage[table]{xcolor}
\usepackage{multirow}
\usepackage{afterpage}

\usepackage[dvipsnames]{xcolor}
\usepackage{pdfpages}
\usepackage{csquotes}
\usepackage{geometry}
\usepackage{lipsum}
\usepackage{soul}
\usepackage[labelfont=bf]{caption}
\biboptions{sort&compress}
\usepackage{graphicx}
\usepackage{amssymb}
\usepackage[symbol]{footmisc}
\usepackage{xcolor}
\usepackage{lineno}
\usepackage{amsmath}
\usepackage{gensymb}
\usepackage{float}
\usepackage{dirtytalk}
\usepackage{adjustbox}
\usepackage{placeins}
\usepackage{natbib}
\usepackage{booktabs}
\definecolor{Blueish}{HTML}{0066C4}
\definecolor{Redish}{HTML}{CC1A1A}
\definecolor{Greenish}{HTML}{529241}

\usepackage[most]{tcolorbox}
\definecolor{notegray}{RGB}{245,245,245} 

\tcbset{
  abbrevbox/.style={
    colback=notegray,
    colframe=black!30,
    boxrule=0.5pt,
    arc=2pt,
    left=8pt, right=8pt, top=6pt, bottom=6pt,
    fontupper=\small, fontlower=\small, fonttitle=\bfseries,
    width=\columnwidth, 
    before skip=1em,    
    after skip=1em,      
  }
}

\journal{}

\begin{document}
\begin{frontmatter}


\title{Real-time nonlinear inversion of magnetic resonance elastography with operator learning}

\author{Juampablo E. Heras Rivera, MSc*,$^{1}$ Caitlin M. Neher, MSc*,$^{1}$ and Mehmet Kurt, PhD$^{1}$}

\vspace{1em}

\address{\it$^{1}$University of Washington, Seattle, WA, United States\\}

\begin{abstract}

\textbf{Purpose:} To develop and evaluate an operator learning framework for nonlinear inversion (NLI) of brain magnetic resonance elastography (MRE) data, which enables real-time inversion of elastograms with comparable spatial accuracy to NLI. \\[1.5pt]

\textbf{Materials and Methods:} In this retrospective study, 3D MRE data from 61 individuals (mean age, 37.4 years; 34 female) were used for development of the framework. A predictive deep operator learning framework (oNLI) was trained using 10-fold cross-validation, with the complex curl of the measured displacement field as inputs and NLI-derived reference elastograms as outputs. A structural prior mechanism, analogous to Soft Prior Regularization in the MRE literature, was incorporated to improve spatial accuracy. Subject-level evaluation metrics included Pearson’s correlation coefficient, absolute relative error, and structural similarity index measure between predicted and reference elastograms across brain regions of different sizes to understand accuracy. Statistical analyses included paired t-tests comparing the proposed oNLI variants to the convolutional neural network baselines. \\[1.5pt]

\textbf{Results:} In comparison to convolutional architectures, oNLI showed superior accuracy when used to reconstruct both the storage and loss moduli ($\mu'$ and $\mu''$, respectively) from complex curl fields. Whole brain absolute percent error was 8.4\% $\pm$ 0.5\% ($\mu'$) and 10.0\% $\pm$ 0.7\% ($\mu''$) for oNLI and 15.8\% $\pm$ 0.8\% ($\mu'$) and 26.1\% $\pm$ 1.1\% ($\mu''$) for CNNs. Additionally, oNLI outperformed convolutional architectures as per Pearson’s correlation coefficient $r$ in the whole brain and across all subregions for both the storage modulus and loss modulus (p < .05). In the thalamus, oNLI achieved a mean $\mu'$ correlation of $r$ = 0.92 versus $r$ = 0.87 for CNNs. In the hippocampus, oNLI achieved $r$ = 0.84 versus $r$ = 0.65. Across white matter, oNLI achieved $r$ = 0.93 versus $r$ = 0.78, and in the cortex, $r$ = 0.92 versus $r$ = 0.75. Finally, oNLI predictions of $\mu''$ significantly outperformed CNNs in the whole brain ($r$ = 0.96 versus 0.82), cerebral cortex ($r$ = 0.95 versus 0.79), white matter ($r$ = 0.96 versus 0.81), thalamus ($r$ = 0.95 versus 0.88), and hippocampus ($r$ = 0.91 versus 0.68). \\[1.5pt]

\textbf{Conclusion:} The oNLI framework enables real-time MRE inversion (30,000x speedup), outperforming CNN-based approaches and maintaining the fine-grained spatial accuracy achievable with NLI in the brain.

\end{abstract}

\end{frontmatter}

\section{Introduction}

Magnetic resonance elastography (MRE) \cite{MRE} is a quantitative method for noninvasively obtaining the mechanical properties of sub-superficial organs of interest, such as the liver, brain, and kidneys \cite{MRE1995, MRE2001, Rouviere2011KidneyMRE}. In the liver, MRE is the most accurate noninvasive technique for diagnosing and staging fibrosis \cite{Singh2015MRE_NAFLD}, and has become a routine clinical scan. In the brain, a growing body of evidence suggests that viscoelasticity derived from MRE is sensitive to aging and disease \cite{Lv2020Aging, Feng2024Neurodegeneration} and regional properties correlate with age \cite{Hiscox2021AgingBrainMRE}, cognitive decline \cite{Pavuluri2025Cognition}, ApoE status \cite{hiscox2025mr}, and perfusion \cite{Neher2025}. 

MRE involves three components: the mechanical actuation of the tissue of interest, a phase-contrast MR pulse sequence to encode three-directional displacements during actuation, and post-processing using an inversion algorithm to recover mechanical properties of the tissue from the displacement data. Frequently reported properties are shear stiffness and damping ratio, which provide information on the structural integrity and viscosity of the tissue of interest.  Numerous inversion algorithms have been developed to relate tissue displacement to mechanical properties, each with different material model assumptions, computational costs, and final property estimates. 

When a linear and locally homogeneous material is assumed, the complex shear modulus, \(\mu\), is defined by the Helmholtz equation: 
\begin{equation}
\label{eqn:helmholtz}
    \mu=\frac{-\rho\omega^2\mathbf{u}}{\nabla^2\mathbf{u}}
\end{equation}
where $\mathbf{u}$ is the complex harmonic displacement vector (or its curl), $\rho$ is the density, and $\omega$ is the frequency of mechanical actuation. Inversion approaches based on this assumption used in practice include direct inversion \cite{Romano1998DI}, local frequency estimation \cite{Manduca1996LWE}, and algebraic inversion of the differential equation \cite{oliphant2001complex}. Applying this material model simplifies the inversion and allows for fast computations on the order of seconds, but leads to artifacts near boundaries and inaccuracies around small features. This limits the applicability of these approaches, as interfaces and small anatomical regions like the sub-cortical gray matter cannot be distinctly resolved.

By removing the local homogeneity assumption, the complex shear modulus is defined by the heterogeneous form of the time-harmonic Navier's equation, which describes the evolution of displacement fields in a viscoelastic medium as: 
\begin{equation}
\label{eqn:navier}
    \nabla \cdot (\mu(\nabla \mathbf{u} +\nabla\mathbf{u}^T)) +\nabla(\lambda\nabla \cdot\textbf{u} ) = -\rho\omega^2 \mathbf{u},
\end{equation}
where $\mathbf{u}$ is the complex harmonic displacement vector, $\mu$ is the complex viscoelastic shear modulus, $\lambda$ is the second Lam\'e parameter, $\rho$ is the density, $\omega$ is the frequency of mechanical actuation. The most common inversion approach used for this material model is nonlinear inversion (NLI), a finite-element-based iterative method that estimates viscoelastic parameters by minimizing the difference between measured and simulated displacement fields \cite{VanHouten2001, mcgarry2012multiresolution}. NLI can characterize the sub-cortical material property changes associated with learning \cite{Heselton2025Putamen}, cognition \cite{Delgorio2023HippocampalSub, Delgorio2022Hippocampal}, and epileptic foci \cite{Huesmann2020Hippocampal}. Additionally, heterogeneity allows for the accurate representation of stiffness at interfaces \cite{Wu2025DBS}. While NLI is better suited for organs with high tissue complexity, and makes fewer assumptions, it is several orders of magnitude slower than direct inversion. The high computational cost of NLI, which requires more than four hours of runtime using 32 CPU cores for a standard $132\times132\times60$ matrix size, makes it difficult to implement clinically when real-time results are desired.

Here, we frame inversion of \eqref{eqn:navier} as done in NLI as a neural operator learning \cite{neuraloperator} problem. We term this method as operator NLI (oNLI), where the goal is to learn a mapping between MRE displacement fields and complex shear stiffness in the infinite-dimensional function space setting. Unlike finite-element based methods which require an iterative optimization from scratch for each subject, oNLI learns a general operator that maps any subject's MRE displacement/curl fields to their respective shear moduli in one forward pass.

\begin{tcolorbox}[abbrevbox,float,floatplacement=htbp]
\footnotesize
{\normalsize\bfseries Abbreviations}\\[2pt]
MRE: magnetic resonance elastography, NLI: nonlinear inversion, oNLI: operator learning framework for NLI, SSIM: structural similarity index measure, CNN: convolutional neural network, SPADE: spatially-adaptive normalization.

\vspace{6pt}
{\normalsize\bfseries Summary}\\[2pt]
The operator learning framework for NLI (oNLI) enables inversion 
in fractions of a second, while maintaining the fine-grained spatial 
accuracy achievable with NLI in the brain.

\vspace{6pt}
{\normalsize\bfseries Key Points}
\begin{itemize}[leftmargin=*, itemsep=4pt]  
    \item oNLI, a deep operator learning framework, was developed for brain MRE nonlinear inversion using curl fields as inputs and NLI elastograms as reference outputs. Additionally, a mechanism for incorporating structural priors was introduced.
    \item Across 61 subjects and 10-fold cross-validation, oNLI achieved higher Pearson’s correlation with ground truth NLI than convolutional neural networks in all major brain subregions (e.g., $\mu'$, white matter: $r$ = 0.93 vs. 0.78; $\mu''$, hippocampus: $r$ = 0.91 vs. 0.68).
    \item oNLI reconstructed elastograms ~30,000× faster than conventional NLI (fractions of a second vs. multiple hours) while maintaining fine-grained spatial accuracy.
\end{itemize}

\vspace{6pt}
{\normalsize\bfseries Keywords}\\[2pt]
Magnetic Resonance Elastography, Operator Learning, Brain, Nonlinear Inversion, 
Deep Learning, SPADE, Convolutional Neural Networks.

\end{tcolorbox}

By framing NLI as an operator learning problem, oNLI provides a resolution-invariant mapping, overcoming the issues of convolutional neural network (CNN) approaches, which fail to generalize beyond one imaging resolution. Although the upfront training cost is high, once deployed oNLI enables real-time nonlinear inversion ($<1$ second on one GPU, $<30$ seconds on one CPU), representing a \textbf{30,000x} speedup over NLI. We train instances of oNLI using an openly available 3D brain MRE dataset, and demonstrate that elastograms produced by oNLI show excellent visual and quantitative agreement with ground-truth elastograms when evaluated on held-out subjects. Finally, we introduce SPADE-oNLI, a variant for the encoding of spatial priors relevant to the inversion process, similar to the Soft Prior Regularization approach introduced in \cite{mcgarry2013softprior}.

\begin{figure*}[t]
  \centering
  \includegraphics[width=\textwidth]{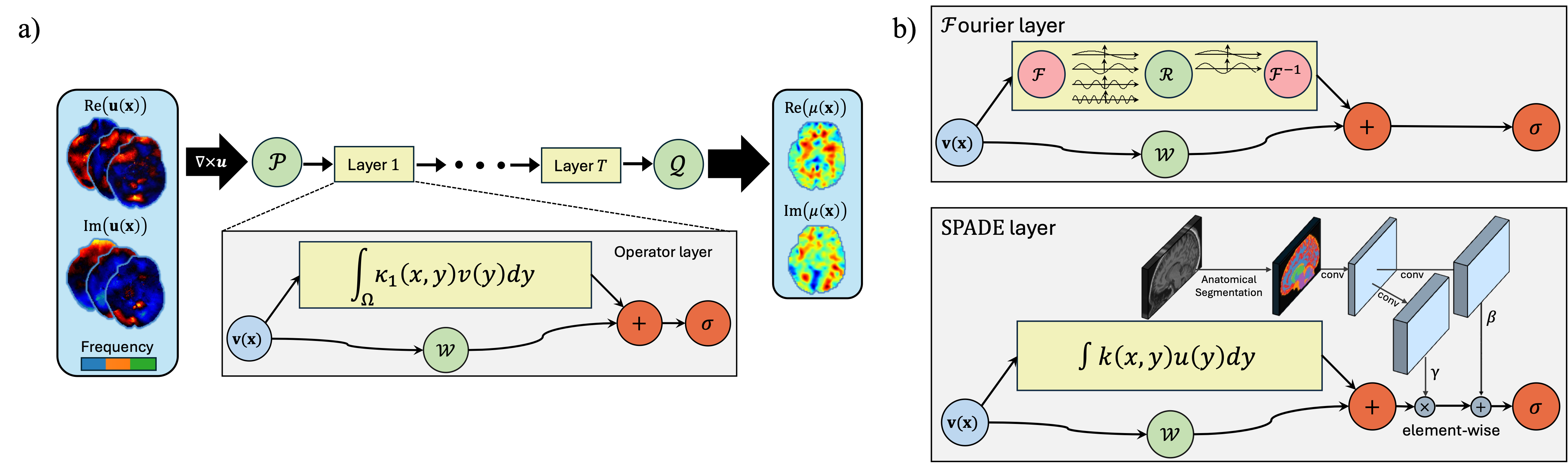}
  \caption{\textbf{a)} oNLI overview: Complex 3D displacement maps, $\mathbf{u}(\mathbf{x})=(u_x, u_y, u_z)\in\mathbb{C}^3$ for $x$ on the imaging domain $x\in\Omega\subset\mathbb{R}^3$, are input to the neural operator, their curl is computed, lifted to a higher dimension by linear mapping $\mathscr{P}$, processed by $T$ Operator layers, and projected to the complex shear modulus, $\mu(\mathbf{x})\in\mathbb{C}$, by the linear map $\mathscr{Q}$. \textbf{b)} Diagrams of operator layers used in this work to instantiate oNLI. Top: Diagram of Fourier Layer introduced in \cite{li2021fourierneuraloperatorparametric}, where the operator kernel is parametrized in the Fourier domain. Bottom: Diagram of the SPADE layer used in the SPADE-oNLI variant. Anatomical T1 scans are processed using SynthSeg \cite{billot2023synthseg} to obtain anatomical segmentations, which are passed through convolutional layers to extract spatially varying statistics, $\gamma$ and $\beta$. These statistics are then used to modulate the affine parameters of instance normalization. }
  \label{fig:onli}
\end{figure*}

\section{Methods}

\subsection{Study Design}

This work is a retrospective study using previously collected data. The primary goal is to design a framework for training deep operators to perform nonlinear inversion of magnetic resonance elastography data in real-time. 

\subsection{Data}

\subsubsection*{Dataset Description}
This study used a publicly available dataset from the Brain Biomechanics Imaging Resources in the Neuroimaging Tools and Resources Collaboratory (NITRC) for training and validation. The specific data used was collected at the University of Delaware, where 61 subjects were scanned on a Siemens Prisma 3T scanner with a 64 channel head and neck coil. Each subject dataset includes an anatomical T1 image at 0.8 mm isotropic resolution and a 3D MRE multiband, multishot spiral sequence aquisition at 30, 50, and 70 Hz at 1.5 mm isotropic resolution. The MRE displacements were generated in the anterior-posterior direction using the Resoundant pneumatic actuation system with a soft pillow driver. High-resolution mechanical property maps were obtained using the NLI algorithm \cite{mcgarry2012multiresolution}. 

\begin{figure}[ht]
    \centering
    \includegraphics[width=0.9\linewidth]{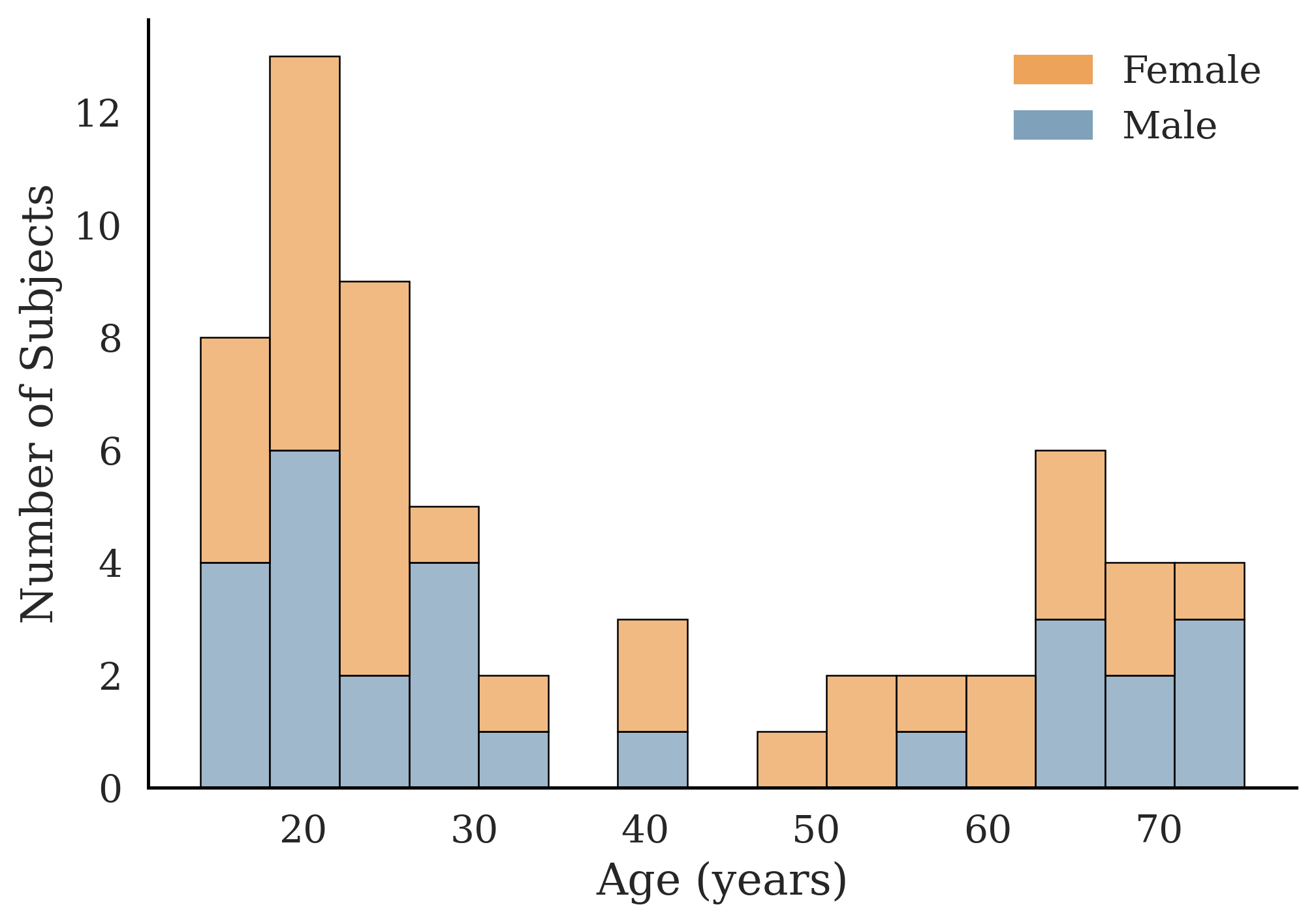}
    \caption{Histogram showing the age distribution of the subjects used for training and validation, including male (blue) and female (orange) partitions for each bin. }
    \label{fig:age_hist}
\end{figure}

The subjects used for training and validation are between 14 to 75 years old, with 34F/27M and an age distribution described by the histogram in Figure \ref{fig:age_hist}. These 61 subjects have a mean age of 37.43 $\pm$ 20.46 years.

For this retrospective study, data were included based on public availability. To our knowledge, all subjects were without neurological conditions at the time the data was collected. 

\subsubsection*{Data Format}
For each subject and actuation frequency, complex-valued displacement fields from MRE acquisitions and their derived curl fields, along with complex shear modulus maps (storage and loss moduli) estimated via NLI were used. Anatomical T1-weighted scans were coregistered to the MRE reference space to enable voxel-wise correspondence to MRE fields.

\subsubsection*{Inputs to oNLI}
For each spatial location $\mathbf{x}$ on the imaging domain $\Omega\subset\mathbb{R}^3$, the complex-valued 
displacement field $\mathbf{u}(\mathbf{x}) \in \mathbb{C}^3$ was obtained from 
the temporal Fast Fourier Transform (FFT) \cite{fft} of the measured displacement $\mathbf{r}(\mathbf{x},t)$, $t\in \mathbb{R}$,
at the actuation frequency $\omega$:
\begin{equation}
\mathbf{u}(\mathbf{x}) 
= \frac{1}{T} \int_{0}^{T} \mathbf{r}(\mathbf{x},t)\, e^{-i \omega t}\, dt.
\end{equation}

The real and imaginary components of $\mathbf{u}(\mathbf{x})$ were each stored as arrays of shape $(160 \times 160 \times 80 \times 3)$, where the first three dimensions index the spatial grid and the last dimension corresponds to the three displacement directions $(x,y,z)$. From this displacement field, the curl $\left(\nabla\times\mathbf{u}(\mathbf{x})\right)$ is obtained, which aids inversion by suppressing longitudinal waves and enhancing the shear wave components produced by mode conversion. For all experiments, the real and imaginary components of the curl fields were concatenated along the fourth dimension, together with an additional channel containing the actuation frequency (in Hz) divided by $100$. This yielded arrays of shape $(160 \times 160 \times 80 \times 7)$, which were used as inputs to the oNLI framework.

\subsubsection*{Outputs from oNLI}
The outputs from oNLI are complex shear modulus fields 
$\mu(\mathbf{x}) = \mu'(\mathbf{x}) + i\,\mu''(\mathbf{x}) \in \mathbb{C}^3$, 
where $\mu'$ and $\mu''$ correspond to the storage and loss moduli, respectively. 
These fields are represented as four-dimensional arrays of shape 
$(160 \times 160 \times 80 \times 2)$, where the first three dimensions index 
the spatial grid and the fourth dimension indexes the real and imaginary components of the complex shear modulus. 






\subsubsection*{Missing Data}

Five NITRC subject datasets were excluded completely due to missing MRE data. Additionally, one subject was missing 30 Hz MRE data and four subjects were missing 70 Hz MRE data. These subjects with partially missing data were not excluded; instead, their available data were used for training and validation.


\subsection{Ground Truth}
The ground truth material property maps were computed using the nonlinear inversion algorithm developed in \cite{mcgarry2012multiresolution}. NLI iteratively solves for the shear modulus as described by the heterogeneous form of the time-harmonic Navier’s equation (Equation \eqref{eqn:navier}).

\subsection{Data Partitions}
To evaluate model architectures and extract performance statistics, 10-fold cross-validation was performed. The dataset contains scans from 61 subjects, each at multiple frequencies, so the folds were divided on a per-subject basis to avoid data leakage between training and validation sets. The dataset was randomly split into 10 folds such that each fold contained 54-55 subjects in the training set and 5-6 in the validation set.

\subsection{Model}



\subsection*{Neural Operators for MRE Inversion}

We introduce \textbf{oNLI} (operator nonlinear inversion), a neural operator framework for MRE. 
Given complex curl fields $\nabla \times \mathbf{u}(\mathbf{x}) \in \mathbb{C}^3$ defined on the imaging 
domain $\Omega \subset \mathbb{R}^3$, the goal is to recover the spatially varying shear modulus 
$\mu(\mathbf{x}) \in \mathbb{C}^3$.

Formally, let $\mathcal{U}$ denote the function space of complex curl fields 
$\nabla \times \mathbf{u}:\Omega \to \mathbb{C}^3$ and $\mathcal{Y}$ the function space of complex modulus fields 
$\mu:\Omega \to \mathbb{C}^3$. The inversion task is expressed as learning the operator 
$\mathcal{G}: \mathcal{U} \to \mathcal{Y}$, defined by
\begin{equation}
    \mathcal{G}(\nabla \times \mathbf{u})(\mathbf{x}) = \mu(\mathbf{x}), 
    \quad \mathbf{x}\in\Omega,
\end{equation}
where $\mathcal{G}$ is resolution-invariant and mesh-free. Unlike finite-dimensional networks that depend on voxel discretization, neural operators approximate mappings between function spaces, allowing a single set of learned parameters to generalize across resolutions and subjects. This is highly beneficial for MRE, where voxel sizes vary across scanners and protocols.


In practice, $\mathcal{G}$ is implemented by composing learnable kernel operators 
$\{\mathcal{K}_i\}_{i=1}^T$ with pointwise nonlinearities, where $T$ denotes the number of kernel layers. Each kernel operator has the form
\begin{equation}
    (\mathcal{K}_i v)(x) = \int_{\Omega} \kappa_i(x,y)\,v(y)\,dy,
\end{equation}
such that
\begin{equation}
    \mathcal{G} = \mathscr{Q} \circ \sigma \circ \mathcal{K}_T 
    \circ \sigma \circ \cdots \circ \sigma \circ \mathcal{K}_1 \circ \mathscr{P},
\end{equation}
where $\sigma$ denotes a pointwise nonlinearity (e.g., ReLU), $v$ denotes an intermediate feature field, and $\mathscr{P}$ and $\mathscr{Q}$ are learned 
linear maps that lift the input curl $\nabla \times \mathbf{u}$ to a higher-dimensional feature space and 
project intermediate representations back to the complex modulus field, respectively. A diagram of the oNLI framework is shown in 
Figure \ref{fig:onli}.


\subsection*{Fourier Neural Operator}
\label{model:fno}

To instantiate oNLI, we adopt the Fourier neural operator (FNO).  FNOs are efficient approximators as they parametrize the operator kernel in the Fourier domain. Given input $v$, a Fourier layer computes
\begin{equation}
    (\mathcal{K}(v))(x) = \mathcal{F}^{-1}\!\left( R \cdot \mathcal{F}(v) \right)(x),
\end{equation}
where $\mathcal{F}$ and $\mathcal{F}^{-1}$ denote the Fourier transform and its inverse, respectively, and $R$ is a linear transform on the lower Fourier modes. By combining global spectral convolution with local nonlinearities, FNOs approximate highly nonlinear PDE-driven operators with quasi-linear complexity \cite{kovachki2021neural}. Importantly, since the operator is learned in continuous space, the same parameters apply across discretizations, allowing for super-resolution and fast inference (millisecond-scale per volume) without retraining or finetuning. 

\subsubsection*{Incorporating anatomical priors}
\begin{figure}[ht]
        \centering
        \includegraphics[width=\linewidth]{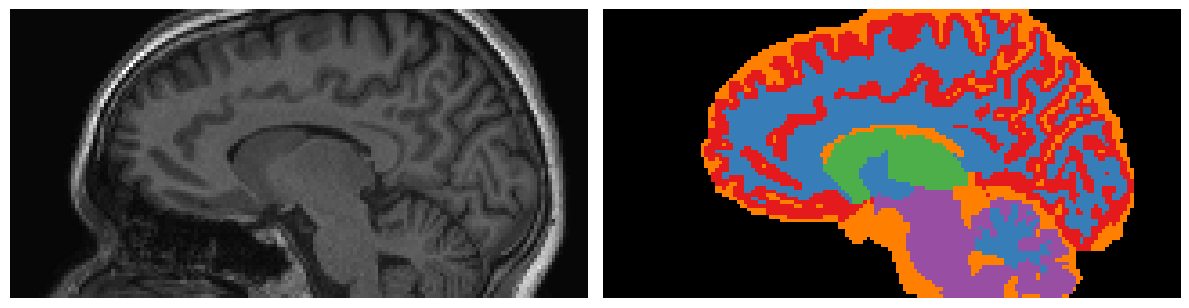}
        \caption{
        T1-weighted MRI (left) and corresponding 6-region segmentation mask (right) generated using SynthSeg. 
        Regions include: background (black), cortical gray matter (red), white matter (blue), subcortical gray matter (green), brainstem/cerebellum (purple), and cerebrospinal fluid (CSF; orange).
        }
        \label{fig:t1-synthseg}
    \end{figure}

T1-weighted MRI acquired alongside MRE provides high-resolution anatomical structure information. We incorporate these priors into oNLI using whole-brain segmentations generated by \texttt{SynthSeg} \cite{billot2023synthseg}, which is substantially faster than \texttt{recon-all} from FreeSurfer \cite{fischl2012freesurfer} while maintaining acceptable accuracy. Segmentations from \texttt{SynthSeg} include over 33 anatomical regions. We group these into six broader categories: cortical gray matter, white matter, and subcortical gray matter, which have statistically distinct material properties \cite{hiscox2020standard}; as well as cerebrospinal fluid (CSF), brainstem/cerebellum, and background, which are typically noisy in MRE data and thus treated as separate regions. Segmentation masks are represented as a one-hot encoded vector, resulting in a 6-channel binary segmentation mask of shape $ \{0, 1\}^{6 \times H \times W \times D} $, where $H$, $W$, and $D$ denote the spatial dimensions. Figure \ref{fig:t1-synthseg} shows a representative sagittal slice of a T1-weighted scan and the corresponding anatomical segmentation mask.

To integrate these anatomical priors into the oNLI framework, we adopt the SPADE \cite{SPADE} approach, a spatially adaptive normalization method originally developed for semantic image synthesis. SPADE modulates the affine parameters of instance normalization using spatially varying statistics derived from the segmentation map. In SPADE-oNLI, the multi-channel segmentation mask \( S \in \{0, 1\}^{6 \times H \times W \times D} \) is processed by a small convolutional network to predict spatially adaptive scale \( \gamma(S)\ \) and bias \(\beta(S) \in \mathbb{R}^{C \times H \times W \times D} \) maps. Given an intermediate feature map \( F \in \mathbb{R}^{C \times H \times W \times D} \), the modulated output is:
    \[
    \text{SPADE}(F, S) = \gamma(S) \cdot \text{IN}(F) + \beta(S)
    \]
    where \( \text{IN}(F) \) is the instance-normalized version of $F$ before affine transformation.
A diagram illustrating a SPADE-oNLI layer is shown in Figure \ref{fig:onli}b). This conditioning is effective in preserving high-frequency features, counteracting the low-frequency bias of neural operators observed in \cite{LowFreqBiasFNO} and imposing structural priors. We observe that this layer-wise conditioning strategy preserves the conditioning throughout the model inference much better than passing the conditioning as an additional input. Unlike the original SPADE paper, which recommends 3x3 convolutions tied to a fixed grid resolution, we use 1x1 convolutions, so that conditioning is independent of spatial discretization, preserving oNLI's resolution invariant property. For the SPADE-oNLI variant considered in the experiments, the operator kernel was parametrized in the Fourier domain as in the Fourier Neural Operator.

\subsection{U-Net}
\label{model:unet}
As a baseline comparison, the 3D U-Net \cite{unet} architecture was used. U-Net has been widely adopted for medical image processing tasks due to its strong inductive bias for spatially localized features.  The model consists of an encoder–decoder structure with skip connections that preserve spatial detail across resolutions. Unlike oNLI, which learns an operator between function spaces, U-Net learns a finite-dimensional function mapping defined on discretized grids. As a result, the model is tied to a fixed discretization, and inversion of inputs with different resolutions requires resampling to match the resolution of the training data. 

\subsubsection*{Model architecture}


 Three model architectures were considered for evaluation of the proposed method, namely: U-Net (baseline), oNLI, and SPADE-oNLI. All models were configured with seven input channels: 6 channels for the real and imaginary curl components in three orthogonal directions, and 1 channel for frequency conditioning. All models had two output channels, corresponding to the real and imaginary parts of the complex shear modulus. For all experiments, the number of model parameters was fixed to between 84–85 million to enable a fair comparison between methods while ensuring compatibility with smaller GPUs and maintaining practical inference times (sub-second on GPU, $<30$ seconds on CPU). 

The baseline U-Net model was configured with 4 encoder levels with channels (84, 168, 336, 672), 3 downsampling steps (stride 2), and residual blocks with instance normalization. 

The oNLI and SPADE-oNLI models were both configured with an FNO parametrization of the operator kernel. However, we emphasize that the oNLI formulation is not limited to Fourier kernels; in principle, oNLI can be implemented with alternative operator kernels, such as wavelets. In both models, each layer used $5$ stacked layers, each with $20$ Fourier modes per spatial dimension and width $23$. The SPADE-oNLI variant used the 3D SPADE block at each layer, with input channels equal to the number of classes in the one-hot anatomical segmentation, a single hidden convolutional layer of 32 channels, and a final convolutional layer with 23 channels to match the FNO layer width. For all convolutional layers, instance normalization and kernel size 1 were used to preserve resolution invariance.

\subsection{Training}
\subsubsection*{Data normalization}
 Inputs and targets were normalized independently for curl and stiffness fields using per-channel unit Gaussian standardization. Specifically, we applied the \texttt{UnitGaussianNormalizer} implementation from the \texttt{neuraloperators} library \cite{kovachki2021neural, kossaifi2024neural}, with the mean and standard deviation estimated from the training set. After forward passes, model predictions were denormalized using the inverse of the preprocessing step, to preserve stable training while still recovering the full variability in magnitude scales across subjects and frequencies.

\subsubsection*{Training Details}
Following the hyperparameter choices in \cite{peng2024fourier}, all models were trained using the Adam optimizer with an initial learning rate of $10^{-3}$ and a weight decay of $10^{-4}$. The cosine annealing learning rate scheduler with $T_{\max}=\text{epochs}\times\lfloor N_{\text{train}}/\text{batch size}\rfloor$ was used. All models were trained for $50$ epochs with a batch size of $1$. 

The relative $L_{2}$ loss was used as an optimization metric, as specified in \cite{kovachki2021neural}. For predictions $\mu^{oNLI} \in \mathbb{R}^{N \times M}$ and ground-truth fields 
$\mu^{\mathrm{NLI}} \in \mathbb{R}^{N \times M}$, where $N$ is the batch size 
and $M$ is the number of voxels per sample, the loss is defined as
\begin{equation}
\label{eq:loss}
\mathcal{L}^{data}_{2}(\mu^{oNLI}, \mu^{\mathrm{NLI}}) 
= \frac{1}{N} \sum_{i=1}^{N} 
\frac{\left\| \mu^{oNLI}_i - \mu^{\mathrm{NLI}}_i \right\|_{2}}
     {\left\| \mu^{\mathrm{NLI}}_i \right\|_{2}} .
\end{equation}
Here, $\| \cdot \|_{2}$ denotes the $L_{2}$-norm. This formulation normalizes 
the prediction error by the target norm to preserve scale invariance across subjects and frequencies.

\subsubsection*{Compute resources and software used}
Each model was trained using 8 CPU cores and a single NVIDIA A40 GPU on the University of Washington's Hyak supercomputing cluster. Model development was done using \texttt{PyTorch} \cite{paszke2019pytorch}, and data analysis with \texttt{Python}.


\begin{table*}[t]
\centering
\caption{Model performance by parameter ($\mu'$ and $\mu''$) and region (whole brain, cerebral cortex, white matter, thalamus, and hippocampus) across pooled validation subjects. Pearson's correlation coefficient ($r$), absolute percent error (APE; \%), and structural similarity index measure (SSIM) are reported. SPADE-oNLI outperformed U-Net in the whole brain and the selected regions except for the SSIM of $\mu'$ for the whole brain, in which U-Net performed the best.}
\label{tab:model_performance}

\setlength{\tabcolsep}{3.5pt}
\renewcommand{\arraystretch}{1.05}
\tiny

\begin{adjustbox}{width=\textwidth}
\begin{tabular}{|c|c|c|c|c|c|c|}


\hline
\multirow{2}{*}{Model} 
& \multicolumn{3}{c|}{\textbf{Storage Modulus ($\mu'$)}} 
& \multicolumn{3}{c|}{\textbf{Loss Modulus ($\mu''$)}} \\
\cline{2-7}
&  Pearson's $r$ ($\uparrow)$ & APE (\%) ($\downarrow$) & SSIM ($\uparrow$) 
& Pearson's $r$ ($\uparrow$) & APE (\%) ($\downarrow$) & SSIM ($\uparrow$) \\
\hline

\rowcolor{gray!15} \multicolumn{7}{|c|}{\textbf{Whole Brain}} \\
\hline

U\text{-}Net      & 0.782 $\pm$ 0.023 & 15.8 $\pm$ 0.8 & \textbf{0.720 $\pm$ 0.043} & 0.822 $\pm$ 0.020 & 26.1 $\pm$ 1.1 & 0.520 $\pm$ 0.041 \\
oNLI              & 0.907 $\pm$ 0.012 & 9.4 $\pm$ 0.6  & 0.670 $\pm$ 0.041 & 0.944 $\pm$ 0.008 & 10.7 $\pm$ 0.7 & 0.595 $\pm$ 0.049 \\
SPADE\text{-}oNLI & \textbf{0.934 $\pm$ 0.008} & \textbf{8.4 $\pm$ 0.5} & 0.679 $\pm$ 0.040 & \textbf{0.956 $\pm$ 0.006} & \textbf{10.0 $\pm$ 0.7} & \textbf{0.598 $\pm$ 0.051} \\
\hline

\rowcolor{gray!15} \multicolumn{7}{|c|}{\textbf{Cerebral Cortex}} \\
\hline

U\text{-}Net      & 0.752 $\pm$ 0.026 & 16.5 $\pm$ 0.8 & 0.731 $\pm$ 0.066 & 0.786 $\pm$ 0.023 & 25.5 $\pm$ 1.1 & 0.510 $\pm$ 0.072 \\
oNLI              & 0.883 $\pm$ 0.015 & 10.6 $\pm$ 0.6 & 0.726 $\pm$ 0.047 & 0.933 $\pm$ 0.010 & 10.8 $\pm$ 0.7 & 0.680 $\pm$ 0.050 \\
SPADE\text{-}oNLI & \textbf{0.915 $\pm$ 0.012} & \textbf{9.6 $\pm$ 0.6} & \textbf{0.735 $\pm$ 0.041} & \textbf{0.946 $\pm$ 0.008} & \textbf{10.8 $\pm$ 0.7} & \textbf{0.685 $\pm$ 0.046} \\
\hline

\rowcolor{gray!15} \multicolumn{7}{|c|}{\textbf{White Matter}} \\
\hline

U\text{-}Net      & 0.775 $\pm$ 0.024 & 15.2 $\pm$ 0.8 & 0.741 $\pm$ 0.067 & 0.806 $\pm$ 0.022 & 28.6 $\pm$ 1.2 & 0.560 $\pm$ 0.042 \\
oNLI              & 0.909 $\pm$ 0.011 & 9.1 $\pm$ 0.5  & 0.755 $\pm$ 0.052 & 0.949 $\pm$ 0.007 & 11.1 $\pm$ 0.7 & 0.711 $\pm$ 0.055 \\
SPADE\text{-}oNLI & \textbf{0.932 $\pm$ 0.010} & \textbf{7.6 $\pm$ 0.6} & \textbf{0.760 $\pm$ 0.051} & \textbf{0.955 $\pm$ 0.007} & \textbf{10.1 $\pm$ 0.8} & \textbf{0.715 $\pm$ 0.054} \\
\hline

\rowcolor{gray!15} \multicolumn{7}{|c|}{\textbf{Thalamus}} \\
\hline

U\text{-}Net      & 0.871 $\pm$ 0.015 & 17.8 $\pm$ 0.9 & 0.438 $\pm$ 0.205 & 0.881 $\pm$ 0.015 & 31.1 $\pm$ 1.8 & 0.203 $\pm$ 0.137 \\
oNLI              & 0.898 $\pm$ 0.012 & 16.1 $\pm$ 1.1 & 0.408 $\pm$ 0.166 & 0.932 $\pm$ 0.009 & 17.3 $\pm$ 1.3 & 0.390 $\pm$ 0.144 \\
SPADE\text{-}oNLI & \textbf{0.915 $\pm$ 0.011} & \textbf{14.1 $\pm$ 1.3} & \textbf{0.442 $\pm$ 0.166} & \textbf{0.946 $\pm$ 0.007} & \textbf{15.6 $\pm$ 1.3} & \textbf{0.403 $\pm$ 0.144} \\
\hline

\rowcolor{gray!15} \multicolumn{7}{|c|}{\textbf{Hippocampus}} \\
\hline

U\text{-}Net      & 0.650 $\pm$ 0.037 & 16.4 $\pm$ 0.9 & 0.734 $\pm$ 0.105 & 0.676 $\pm$ 0.036 & 25.5 $\pm$ 1.3 & 0.629 $\pm$ 0.094 \\
oNLI              & 0.833 $\pm$ 0.020 & 11.3 $\pm$ 0.6 & 0.732 $\pm$ 0.081 & 0.893 $\pm$ 0.019 & 14.3 $\pm$ 1.0 & 0.721 $\pm$ 0.085 \\
SPADE\text{-}oNLI & \textbf{0.841 $\pm$ 0.018} & \textbf{10.6 $\pm$ 0.7} & \textbf{0.754 $\pm$ 0.086} & \textbf{0.910 $\pm$ 0.013} & \textbf{11.8 $\pm$ 0.8} & \textbf{0.734 $\pm$ 0.083} \\
\hline

\end{tabular}
\end{adjustbox}
\end{table*}
 
\begin{table*}[!b]
\centering
\caption{Mean validation relative $L_{2}$ loss (Equation \eqref{eq:loss}) per fold, with across-fold mean, standard deviation (STD), and 95\% confidence interval (CI).}
\label{tab:mean_loss_by_fold}
\renewcommand{\arraystretch}{1.0}
\begin{adjustbox}{width=\textwidth}
\begin{tabular}{|c|cccccccccc|c|c|c|}
\hline
Model & Fold 1 & Fold 2 & Fold 3 & Fold 4 & Fold 5 & Fold 6 & Fold 7 & Fold 8 & Fold 9 & Fold 10 & Mean & STD & 95\% CI \\
\hline
U-Net      & 0.337 & 0.332 & 0.325 & 0.330 & 0.336 & 0.351 & 0.347 & 0.344 & 0.329 & 0.344 & 0.338 & 0.008 & [0.332, 0.344] \\
oNLI       & 0.280 & 0.321 & 0.291 & 0.305 & 0.282 & 0.285 & 0.314 & 0.297 & 0.289 & 0.316 & 0.298 & 0.014 & [0.289, 0.307] \\
SPADE-oNLI & 0.270 & 0.293 & 0.291 & 0.289 & 0.288 & 0.289 & 0.298 & 0.293 & 0.281 & 0.319 & \textbf{0.291} & \textbf{0.012} & \textbf{[0.283, 0.299]} \\
\hline
\end{tabular}
\end{adjustbox}
\end{table*}

\subsection{Evaluation}

In this study, the  predictions from three candidate models: U-Net (baseline), oNLI, and SPADE-oNLI were compared to the NLI ground truth reference. Model evaluation was performed using the held-out datasets from each fold in the 10-fold cross-validation. Predictions from all the held-out subjects across all folds were concatenated to form a pooled validation set, which was used to calculate group-level evaluation metrics. Analyses were conducted on individual brain subregions including: the cerebral cortex, white matter, thalamus, and hippocampus. These regions are commonly studied using NLI and vary significantly in size; here they are used to provide an estimate of model performance on small and large regions. The group-level evaluation metrics reported include: the correlation coefficient between predicted and ground truth values, the absolute relative error, and the structural similarity index (SSIM) between the predicted and corresponding ground truth shear modulus elastograms.  All group-level metrics are reported in Table \ref{tab:model_performance}, and analyzed in the Section \ref{results}.
In addition to the group-level evaluation metrics, we report the fold-level statistics, including the mean, standard deviation, and 95\% confidence interval of the validation relative $L_{2}$ loss across folds. 

\begin{figure*}[!t]
  \centering
  \includegraphics[width=\textwidth]{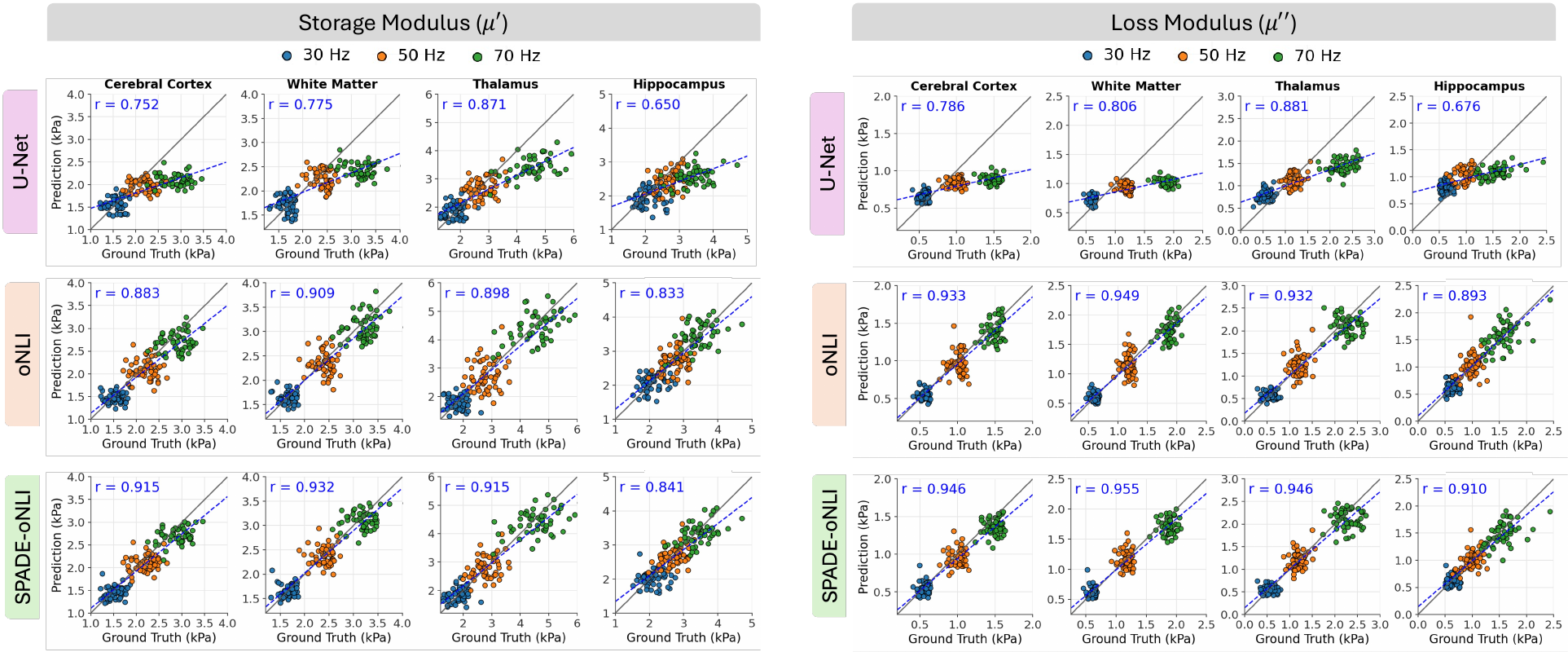}
  \caption{Ground truth vs. predicted mean storage and loss moduli ($\mu'$ and $\mu''$, respectively) across validation subjects ($n=56$) at three actuation frequencies (30, 50, and 70 Hz). Pearson's r with respect to a linear regression fit is reported for each comparison. Across the regions of interest (cerebral cortex, white matter, thalamus, and hippocampus), SPADE-oNLI performs the best with respect to Pearson's r.}
  \label{fig:yequalsx_combined}
\end{figure*}

\begin{figure*}[!b]
  \centering
  \includegraphics[width=\textwidth]{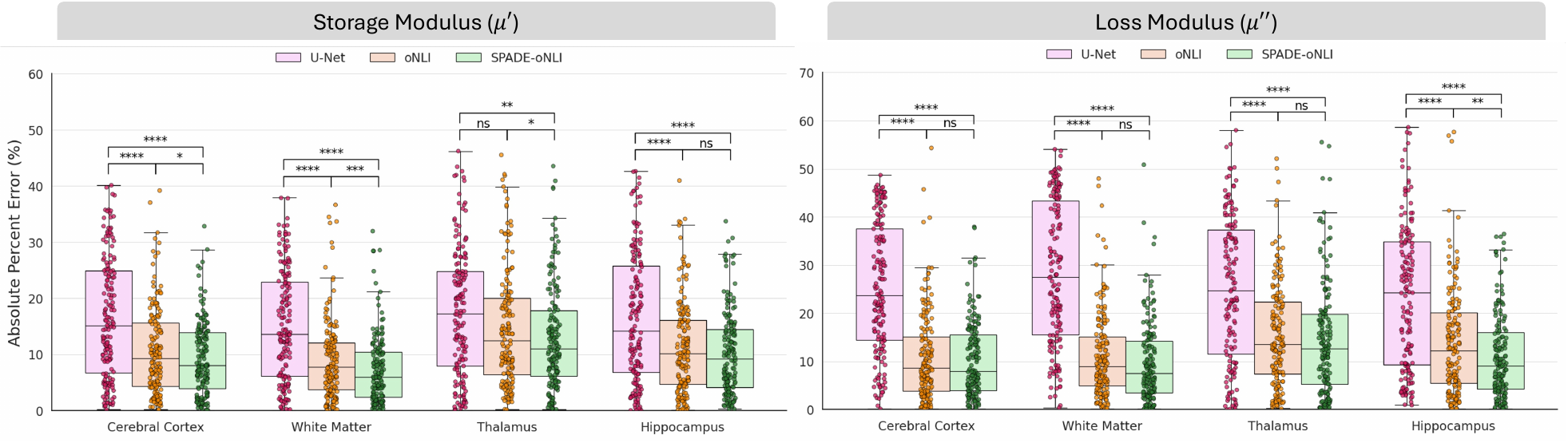}
  \caption{Absolute percent error of the storage and loss moduli ($\mu'$ and $\mu''$, respectively) as evaluated against ground truth mean values for the cerebral cortex, white matter, thalamus and hippocampus. oNLI shows a significant error reduction from U-Net across all regions and both moduli except for the storage modulus in the thalamus, and SPADE-oNLI shows a significant error reduction from U-Net across all cases. ns: p $\geq$ 0.5; *: p $<$ 0.05; **: p $<$ 0.01; ***: p $<$ 0.001; ****; p $<$ 0.0001.}
  \label{fig:percenterror_combined}
\end{figure*}

\begin{figure*}[t]
  \centering
  \includegraphics[width=0.9\textwidth]{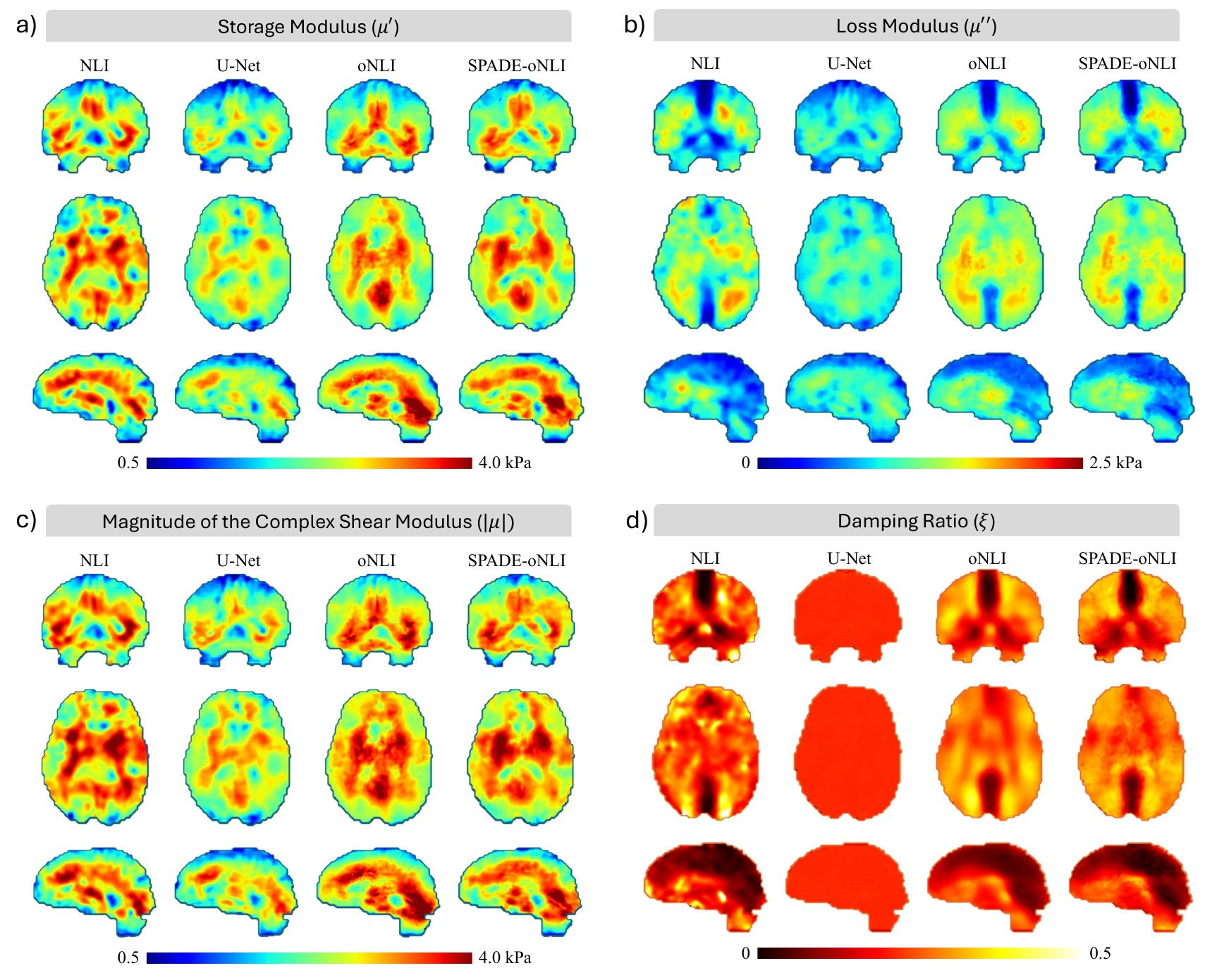}
  \caption{\textbf{a)} Comparison of ground truth and predicted storage modulus ($\mu'$), \textbf{b)} loss modulus ($\mu''$), \textbf{c)} magnitude of the complex shear modulus ($|\mu| = \sqrt{(\mu')^2+(\mu'')^2}$) and \textbf{d)} damping ratio ($\xi = \mu''/2\mu'$) for one representative validation subject at 50Hz. Within each \textbf{a-d)}, from left to right, columns show the respective parameter derived from NLI (ground truth), U-Net prediction, oNLI prediction, and SPADE-oNLI prediction, respectively. From top to bottom: coronal, axial, and sagittal slices are shown for each \textbf{a-d)}.}
  \label{fig:elastograms_combined}
\end{figure*}

\section{Results}
\label{results}




\subsection{Population Demographics}
A total of 61 subjects (34 female and 27 male; mean age, 37.43 years, standard deviation 20.46 years) were included in the model development and evaluation dataset. Among these participants, all were without neurological conditions at the time the data was collected. In the following subsections, fold-wise and group-wise metrics are reported. For group-wise metrics, the predictions for the held-out datasets from all trained models in the 10-fold cross validation are compiled into a larger validation dataset, hereafter referred to as the \enquote{pooled validation set}.

\subsection{Cross-Validation Performance}
The validation relative $L_{2}$ loss (Equation \eqref{eq:loss}) for each of the folds ($n=10$) of all models is shown in Table \ref{tab:mean_loss_by_fold}. The average loss and standard deviation for all folds are 0.338 $\pm$ 0.008, 0.298 $\pm$ 0.014, and 0.291 $\pm$ 0.012 for U-Net, oNLI, and SPADE-oNLI, respectively. Paired t-tests across folds showed that both oNLI $\left(p \approx 3.7 \times 10^{-5}\right)$ and SPADE-oNLI $\left(p \approx 9.0 \times 10^{-7}\right)$ achieved significantly lower validation loss than U-Net, while the difference between oNLI and SPADE-oNLI was not statistically significant $\left(p \approx 0.075\right)$.  The oNLI variants both outperform the U-Net model across all individual folds, indicating a robust improvement. The low average standard deviation indicates similar performance across folds for the full dataset with no outlier folds. Finally, the 95\% confidence intervals for both oNLI variants are  lower than that of the U-Net, indicating with high certainty that performance has improved.

\subsection{Prediction vs. Ground Truth Across Frequencies}
Figure \ref{fig:yequalsx_combined} shows a comparison between ground truth and predicted values for the storage and loss moduli ($\mu'$ and $\mu''$, respectively) across validation subjects in the pooled validation set at three frequencies. Each subplot shows scatter plots of predicted vs. ground truth values for individual regions of interest (cerebral cortex, white matter, thalamus, and hippocampus), with a blue dashed regression line and corresponding r value to indicate the correlation of the model’s predictions with its trend line. Additionally, the $y=x$ reference ideal line is shown for visual comparison, and the correlation coefficients between the model predictions and the ground truth are shown in Table \ref{tab:model_performance}.

For the U-Net variants, we observe that at 30 Hz, U-Net is able to produce estimates which are near the $y=x$ reference line; however, at 50 and 70 Hz, U-Net systematically underpredicts the mean stiffness values across all subregions, even though frequency was included as a conditioning input.  Furthermore, U-Net struggles to accurately model the loss modulus, and predicts near constant mean values across frequencies, indicating that this model architecture is unable to capture the monotonically increasing stiffness-frequency relationship.

For the baseline oNLI variant and the SPADE-oNLI variant we observe that the model predictions closely approach the $y=x$ reference line across all frequencies. Both variants accurately capture the increasing stiffness-frequency relationship in both the storage and loss moduli. Quantitatively, values of $r$ for SPADE-oNLI outperform U-Net in all regions analyzed for $\mu'$ (whole brain: 0.93 versus 0.78; cerebral cortex: 0.92 versus 0.75; white matter: 0.93 versus 0.78; thalamus: 0.92 versus 0.87; and hippocampus: 0.84 versus 0.65) and $\mu''$ (whole brain: 0.96 versus 0.82; cerebral cortex: 0.95 versus 0.79; white matter: 0.96 versus 0.81; thalamus: 0.95 versus 0.88; and hippocampus: 0.91 versus 0.68). 


\subsection{Regional Relative Error Analysis}
Figure \ref{fig:percenterror_combined} shows a comparison of the absolute percent relative errors (APE) for all subjects in the pooled validation set across brain subregions for all considered models. This measure provides an estimate for the error that can be expected from using each model to predict the mean value of the storage and loss moduli in each of the subregions. All APE values are presented in Table \ref{tab:model_performance} as well. Paired t-tests were used to perform pairwise comparisons between models for each region, and significance is reported in Figure \ref{fig:percenterror_combined}. Across the cerebral cortex, white matter, and hippocampus regions, for both storage and loss modulus, we observe that the oNLI variants significantly outperform the U-Net variants with \textbf{$p < 0.0001$}. For the storage modulus in the thalamus, we see that the baseline oNLI variant is not significantly better than the U-Net, but SPADE-oNLI is significantly better than both ($p < 0.05$ vs. baseline oNLI and $p < 0.01$ vs. U-Net). Generally, we observe that SPADE-oNLI significantly outperforms U-Net on both small and large brain regions.

\subsection{Structural Similarity Index Measure (SSIM)}
The SSIM metric was used to further quantify each model's full-brain and regional performance. Whole brain analysis showed that U-Net performed the best for predictions of $\mu'$ with a SSIM of 0.72. However, SPADE-oNLI achieved the best SSIM for the rest of the $\mu'$ predictions (cerebral cortex, white matter, thalamus and hippocampus) and all of the $\mu''$ predictions, including the whole brain. Interestingly, the U-Net predictions of the storage modulus were comparable to SPADE-oNLI across all subregions, but the loss modulus predictions were significantly worse. Since SSIM is a metric which combines luminance, contrast and structure, it provides a more holistic understanding of the spatial distribution of values. From the elastograms in Figure \ref{fig:elastograms_combined}, we see that U-Net correctly predicts the relative storage modulus in the ventricles, white matter, and gray matter. However, it predicts the same spatial distribution at a different scale for the loss modulus, failing to learn a pattern distinct from the storage modulus. This effect is highlighted when the damping ratio is plotted in Figure \ref{fig:elastograms_combined}d), where the voxel-wise calculation $\mu''/2\mu'$ results in an almost completely constant and incorrect elastogram. This occurs because CNNs use the same learned filters across channels, limiting their ability to predict significantly different moduli at different channels.





\section{Discussion}



This study introduces oNLI, a data-driven operator approach to perform MRE inversion in fractions of a second in heterogeneous materials with an accuracy comparable to nonlinear inversion.  We develop this approach using 3D brain MRE data and show that operator learning is significantly more data-efficient during training than convolutional network-based approaches such as U-Net for this application, achieving superior accuracy at multiple actuation frequencies when trained on a dataset of only 61 subjects.  Furthermore, we introduce SPADE-oNLI, which allows for the incroporation of spatial priors during inference, further improving the predictions made by oNLI. The proposed inversion requires minimal computational overhead once trained, and results in a 30,000x speedup over traditional NLI methods. All presented approaches require no additional scans or scan time, which aid in seamless integration for future iterations.

By developing a framework which directly predicts heterogeneous material properties, this work represents progress towards real-time high-fidelity MRE in a clinical diagnostic setting and broadens the diagnostic capability of rapid inversion beyond the liver to more complex organs such as the brain.  This reduced inversion time has the potential to broaden the use of MRE in both research and clinical settings, accelerating the development of NLI-based biomarkers. When trained on a dataset with various resolutions, we hypothesize that the oNLI approach will further outperform U-Net variants, given the continuous nature of the representation learned.

Our study had several limitations, which should be addressed in future works. The proposed model was trained and evaluated using retrospective data from a single site, so further analysis is required to validate the results in prospective multicenter studies. Additionally, training and evaluation using multi-resolution data should be done to further utilize the capabilities of resolution-invariant operator learning models. Furthermore, the methods presented were only evaluated with a healthy cohort, and should be further developed to ensure robust performance in characterizing pathological tissue properties.

In conclusion, this study proposes a data-driven operator learning approach for real-time inversion of Navier's equation with the heterogeneous material assumptions, as done with NLI. We introduce a method to introduce anatomical priors into the model training by using SPADE layers, a concept typically used for semantic image generation. By leveraging anatomical priors, the proposed approach is able to closely match the ground truth NLI predictions with reduced variance and improved accuracy in comparison to U-Net and the FNO-based oNLI. The presented models robustly and signifcantly outperform U-Net with respect to Pearson's r, the APE, and the SSIM as evidenced by 10-fold cross-validation. Future work will primarily focus on a) collecting a larger and more diverse dataset for training and evaluation of the proposed method on multiple sites and resolutions, b) evaluating the proposed approach on clinical data, and c) incorporating physics-informed simulation data to supplement the limited training dataset size.



\section{Author Contributions}
J.E.H.R.: conceptualization, data curation, formal analysis, investigation, methodology, software, validation, writing--original draft, writing--review and editing;
C.M.N.: conceptualization, data curation, formal analysis, investigation, methodology, software, validation, writing--original draft, writing--review and editing;  M.K.: funding acquisition, supervision, project administration, resources, writing--review.

\section{Acknowledgments} 
The authors thank Dr. Curtis Johnson's group for providing the NITRC MRE dataset and 
Dr.\ Matt McGarry for sharing the NLI implementation.

\section{Funding Statement}
The work of Juampablo Heras Rivera was partially supported by the U.S. Department of Energy Computational Science Graduate Fellowship under Award Number DE-SC0024386. The work of Caitlin Neher was partially supported by the National Science Foundation under Award Number NSF CMMI 1953323. 



\begingroup
\renewcommand{\baselinestretch}{0.8}  
\setlength{\bibsep}{0pt plus 0.3ex}   
\small                            
\bibliography{references}

\begin{thebibliography}{10}

\bibitem{MRE}
R.~Muthupillai, D.~J. Lomas, P.~J. Rossman, J.~F. Greenleaf, A.~Manduca, and R.~L. Ehman.
\newblock Magnetic resonance elastography by direct visualization of propagating acoustic strain waves.
\newblock {\em Science}, 269(5232):1854--1857, 1995.

\bibitem{MRE1995}
Raja Muthupillai, David~J. Lomas, Philip~J. Rossman, James~F. Greenleaf, Armando Manduca, and Richard~L. Ehman.
\newblock Magnetic resonance elastography by direct visualization of propagating acoustic strain waves.
\newblock {\em Science}, 269(5232):1854--1857, 1995.

\bibitem{MRE2001}
Armando Manduca, T.~E. Oliphant, M.~A. Dresner, J.~L. Mahowald, S.~A. Kruse, E.~Amromin, J.~P. Felmlee, J.~F. Greenleaf, and R.~L. Ehman.
\newblock Magnetic resonance elastography: Non-invasive mapping of tissue elasticity.
\newblock {\em Medical Image Analysis}, 5(4):237--254, December 2001.

\bibitem{Rouviere2011KidneyMRE}
Olivier Rouvi\`ere, R\'emi Souchon, Ga\"ele Pagnoux, Jean‑Michel M\'enager, and Jean‑Yves Chapelon.
\newblock Magnetic resonance elastography of the kidneys: feasibility and reproducibility in young healthy adults.
\newblock {\em Journal of Magnetic Resonance Imaging}, 34(4):880--886, oct 2011.

\bibitem{Singh2015MRE_NAFLD}
Siddharth Singh, Sudhakar~K. Venkatesh, Zhen Wang, Frank~H. Miller, Utaroh Motosugi, Yoshito Kamiya, and et~al.
\newblock Magnetic resonance elastography for staging liver fibrosis in nonalcoholic fatty liver disease: A pooled individual participant data analysis.
\newblock {\em Hepatology}, 62(3):748--756, 2015.

\bibitem{Lv2020Aging}
Han Lv, Mehmet Kurt, Nan Zeng, Erdal Ozkaya, Fabien Marcuz, Lyndia Wu, Kaveh Laksari, David~B. Camarillo, Kim Butts~Pauly, Zhenchang Wang, and Max Wintermark.
\newblock {MR} elastography frequency–dependent and independent parameters demonstrate accelerated decrease of brain stiffness in elder subjects.
\newblock {\em European Radiology}, 30(12):6614--6623, 2020.

\bibitem{Feng2024Neurodegeneration}
Yuan Feng, Matthew~C. Murphy, Emi Hojo, Fei Li, and Neil Roberts.
\newblock Magnetic resonance elastography in the study of neurodegenerative diseases.
\newblock {\em Journal of Magnetic Resonance Imaging}, 59(1):82--96, 2024.

\bibitem{Hiscox2021AgingBrainMRE}
Lucy~V. Hiscox, Hillary Schwarb, Matthew D.~J. McGarry, and Curtis~L. Johnson.
\newblock Aging brain mechanics: Progress and promise of magnetic resonance elastography.
\newblock {\em NeuroImage}, 232:117889, 2021.

\bibitem{Pavuluri2025Cognition}
KowsalyaDevi Pavuluri, John Huston, Richard~L. Ehman, Armando Manduca, Prashanthi Vemuri, Clifford~R. Jack, Matthew~L. Senjem, and Matthew~C. Murphy.
\newblock Brain mechanical properties predict longitudinal cognitive change in aging and alzheimer's disease.
\newblock {\em Neurobiology of Aging}, 147:203--212, 2025.

\bibitem{hiscox2025mr}
Lucy~V Hiscox, Bradley Karat, Louisa~E Wood, Robert~C Davis, Anne Corbett, Claudia Metzler-Baddeley, Curtis~L Johnson, and Derek~K Jones.
\newblock {MR} elastography reveals lower hippocampal stiffness in middle-aged apoe $\varepsilon$4 carriers without cognitive impairment.
\newblock 2025.

\bibitem{Neher2025}
Caitlin~Maria Neher, Em~Triolo, Fargol Rezayaraghi, Oleksandr Khegai, Priti Balchandani, Matthew McGarry, and Mehmet Kurt.
\newblock Perfusion–mechanics coupling of the hippocampus.
\newblock {\em Interface Focus}, 15(1):20240051, April 2025.
\newblock Published April 4, 2025.

\bibitem{Romano1998DI}
A.~J. Romano, J.~J. Shirron, and J.~A. Bucaro.
\newblock On the noninvasive determination of material parameters from a knowledge of elastic displacements theory and numerical simulation.
\newblock {\em IEEE Transactions on Ultrasonics, Ferroelectrics, and Frequency Control}, 45(3):751--759, May 1998.

\bibitem{Manduca1996LWE}
A.~Manduca, R.~Muthupillai, P.~J. Rossman, J.~F. Greenleaf, and R.~L. Ehman.
\newblock Local wavelength estimation for magnetic resonance elastography.
\newblock In {\em Proceedings of 3rd IEEE International Conference on Image Processing}, volume~3, pages 527--530, Lausanne, Switzerland, 1996.

\bibitem{oliphant2001complex}
Travis~E Oliphant, Armando Manduca, Richard~L Ehman, and James~F Greenleaf.
\newblock Complex-valued stiffness reconstruction for magnetic resonance elastography by algebraic inversion of the differential equation.
\newblock {\em Magnetic Resonance in Medicine: An Official Journal of the International Society for Magnetic Resonance in Medicine}, 45(2):299--310, 2001.

\bibitem{VanHouten2001}
E.~E.~W. Van~Houten, M.~I. Miga, J.~B. Weaver, F.~E. Kennedy, and K.~D. Paulsen.
\newblock Three-dimensional subzone-based reconstruction algorithm for {MR} elastography.
\newblock {\em Magnetic Resonance in Medicine}, 45:827--837, 2001.

\bibitem{mcgarry2012multiresolution}
Matthew~D McGarry, Elijah~E Van~Houten, Carter~L Johnson, John~G Georgiadis, Bradley~P Sutton, John~B Weaver, and Keith~D Paulsen.
\newblock Multiresolution {MR} elastography using nonlinear inversion.
\newblock {\em Medical Physics}, 39(10):6388--6396, October 2012.

\bibitem{Heselton2025Putamen}
H.~J. Heselton, A.~T. Anderson, C.~L. Johnson, N.~J. Cohen, B.~P. Sutton, and H.~Schwarb.
\newblock Putamen stiffness declines with age and is associated with implicit sequence learning outcomes.
\newblock {\em Brain Sciences}, 15(9):947, 2025.

\bibitem{Delgorio2023HippocampalSub}
Peyton~L. Delgorio, Lucy~V. Hiscox, Grace McIlvain, Mary~K. Kramer, Alexa~M. Diano, Kyra~E. Twohy, Alexis~A. Merritt, Matthew D.~J. McGarry, Hillary Schwarb, Ana~M. Daugherty, James~M. Ellison, Alyssa~M. Lanzi, Matthew~L. Cohen, Christopher~R. Martens, and Curtis~L. Johnson.
\newblock Hippocampal subfield viscoelasticity in amnestic mild cognitive impairment evaluated with {MR} elastography.
\newblock {\em NeuroImage: Clinical}, 37:103327, 2023.

\bibitem{Delgorio2022Hippocampal}
Peyton~L. Delgorio, Lucy~V. Hiscox, Ana~M. Daugherty, Faria Sanjana, Grace McIlvain, Ryan~T. Pohlig, Matthew D.~J. McGarry, Christopher~R. Martens, Hillary Schwarb, and Curtis~L. Johnson.
\newblock Structure--function dissociations of human hippocampal subfield stiffness and memory performance.
\newblock {\em Journal of Neuroscience}, 42(42):7957--7968, 2022.

\bibitem{Huesmann2020Hippocampal}
Graham~R. Huesmann, Hillary Schwarb, Daniel~R. Smith, Ryan~T. Pohlig, Aaron~T. Anderson, Matthew D.~J. McGarry, Keith~D. Paulsen, Tracey~Mencio Wszalek, Bradley~P. Sutton, and Curtis~L. Johnson.
\newblock Hippocampal stiffness in mesial temporal lobe epilepsy measured with {MR} elastography: Preliminary comparison with healthy participants.
\newblock {\em NeuroImage: Clinical}, 27:102313, 2020.

\bibitem{Wu2025DBS}
Chengyuan Wu, Mahdi Alizadeh, Mary~K. Kramer, Matthew~B. Kroen, Robert Ziechmann, Feroze~B. Mohamed, Qianhong Wu, and Curtis~L. Johnson.
\newblock Deep brain stimulation electrode deviations are associated with brain stiffness interfaces measured by magnetic resonance elastography.
\newblock {\em Operative Neurosurgery}, page ons.0000000000001523, 2025.
\newblock Published online February 20, 2025.

\bibitem{neuraloperator}
Nikola Kovachki, Zongyi Li, Burigede Liu, Kamyar Azizzadenesheli, Kaushik Bhattacharya, Andrew Stuart, and Anima Anandkumar.
\newblock Neural operator: Learning maps between function spaces with applications to pdes.
\newblock {\em Journal of Machine Learning Research}, 24(89):1--97, 2023.

\bibitem{mcgarry2013softprior}
Matthew McGarry, Courtney~L. Johnson, Bradley~P. Sutton, Elijah~E. Van~Houten, John~G. Georgiadis, John~B. Weaver, and Keith~D. Paulsen.
\newblock Including spatial information in nonlinear inversion {MR} elastography using soft prior regularization.
\newblock {\em IEEE Transactions on Medical Imaging}, 32(10):1901--1909, Oct 2013.
\newblock Epub 2013 Jun 17.

\bibitem{li2021fourierneuraloperatorparametric}
Zongyi Li, Nikola Kovachki, Kamyar Azizzadenesheli, Burigede Liu, Kaushik Bhattacharya, Andrew Stuart, and Anima Anandkumar.
\newblock Fourier neural operator for parametric partial differential equations, 2021.

\bibitem{billot2023synthseg}
Benjamin Billot, Douglas~N. Greve, Oula Puonti, Axel Thielscher, Koen Van~Leemput, Bruce Fischl, Adrian~V. Dalca, and Juan~Eugenio Iglesias.
\newblock {SynthSeg}: Segmentation of brain mri scans of any contrast and resolution without retraining.
\newblock {\em Medical Image Analysis}, 86:102789, 2023.

\bibitem{fft}
James~W Cooley and John~W Tukey.
\newblock An algorithm for the machine calculation of complex fourier series.
\newblock {\em Mathematics of computation}, 19(90):297--301, 1965.

\bibitem{kovachki2021neural}
Nikola~B. Kovachki, Zongyi Li, Burigede Liu, Kamyar Azizzadenesheli, Kaushik Bhattacharya, Andrew~M. Stuart, and Anima Anandkumar.
\newblock Neural operator: Learning maps between function spaces.
\newblock {\em CoRR}, abs/2108.08481, 2021.

\bibitem{fischl2012freesurfer}
Bruce Fischl.
\newblock Freesurfer.
\newblock {\em NeuroImage}, 62(2):774--781, 2012.

\bibitem{hiscox2020standard}
Lucy~V Hiscox, Matthew~DJ McGarry, Hillary Schwarb, Elijah~EW Van~Houten, Ryan~T Pohlig, Neil Roberts, Graham~R Huesmann, Agnieszka~Z Burzynska, Bradley~P Sutton, Charles~H Hillman, et~al.
\newblock Standard-space atlas of the viscoelastic properties of the human brain.
\newblock {\em Human brain mapping}, 41(18):5282--5300, 2020.

\bibitem{SPADE}
Taesung Park, Ming-Yu Liu, Ting-Chun Wang, and Jun-Yan Zhu.
\newblock Semantic image synthesis with spatially-adaptive normalization.
\newblock In {\em Proceedings of the IEEE/CVF conference on computer vision and pattern recognition}, pages 2337--2346, 2019.

\bibitem{LowFreqBiasFNO}
Shaoxiang Qin, Fuyuan Lyu, Wenhui Peng, Dingyang Geng, Ju~Wang, Xing Tang, Sylvie Leroyer, Naiping Gao, Xue Liu, and Liangzhu~Leon Wang.
\newblock Toward a better understanding of fourier neural operators from a spectral perspective.
\newblock {\em arXiv preprint arXiv:2404.07200}, 2024.

\bibitem{unet}
Olaf Ronneberger, Philipp Fischer, and Thomas Brox.
\newblock U-net: Convolutional networks for biomedical image segmentation.
\newblock {\em CoRR}, abs/1505.04597, 2015.

\bibitem{kossaifi2024neural}
Jean Kossaifi, Nikola Kovachki, Zongyi Li, Davit Pitt, Miguel Liu-Schiaffini, Robert~Joseph George, Boris Bonev, Kamyar Azizzadenesheli, Julius Berner, and Anima Anandkumar.
\newblock A library for learning neural operators, 2024.

\bibitem{peng2024fourier}
Wenhui Peng, Shaoxiang Qin, Senwen Yang, Jianchun Wang, Xue Liu, and Liangzhu~(Leon) Wang.
\newblock Fourier neural operator for real-time simulation of {3D} dynamic urban microclimate.
\newblock {\em Building and Environment}, 248:111063, 2024.

\bibitem{paszke2019pytorch}
Adam Paszke, Sam Gross, Francisco Massa, Adam Lerer, James Bradbury, Gregory Chanan, Trevor Killeen, Zeming Lin, Natalia Gimelshein, Luca Antiga, et~al.
\newblock Pytorch: An imperative style, high-performance deep learning library.
\newblock {\em Advances in neural information processing systems}, 32, 2019.

\end{thebibliography}
\endgroup


\end{document}